\documentclass{aa}

\usepackage{graphicx}
\usepackage{txfonts}
\usepackage{multirow}
\usepackage{longtable}

\begin{document}

   \title{Boost recall in quasi-stellar object (QSO) selection from highly imbalanced photometric datasets\thanks{Table \ref {tab:newqso} is only available in electronic form at the CDS via anonymous ftp to cdsarc.u-strasbg.fr (130.79.128.5) or via http://cdsweb.u-strasbg.fr/cgi-bin/qcat?J/A+A/}}

   \subtitle{The reverse selection method}

   \author{
   Giorgio Calderone\inst{1}\fnmsep\thanks{E-mail: giorgio.calderone@inaf.it}\and
    Francesco Guarneri\inst{1,2}\and
    Matteo Porru\inst{1}\and
    Stefano Cristiani\inst{1,3,4}\and
    Andrea Grazian\inst{5}\and
    Luciano Nicastro\inst{6}\and
    Manuela Bischetti\inst{1}\and
    Konstantina Boutsia\inst{7,8}\and
    Guido Cupani\inst{1,3}\and
    Valentina D'Odorico\inst{1,3,9}\and
    Chiara Feruglio\inst{1}\and
    Fabio Fontanot\inst{1,3}
    }

   \institute{INAF--Osservatorio Astronomico di Trieste, Via G.B. Tiepolo, 11, I-34143 Trieste, Italy
     \and
     Dipartimento di Fisica, Sezione di Astronomia, Università di Trieste, via G.B. Tiepolo 11, I-34143, Trieste, Italy
     \and
     IFPU--Institute for Fundamental Physics of the Universe, via Beirut 2, I-34151 Trieste, Italy
     \and
     INFN--National Institute for Nuclear Physics, via Valerio 2, I-34127 Trieste, Italy
     \and
     INAF--Osservatorio Astronomico di Padova, Vicolo dell'Osservatorio 5, I-35122 Padova, Italy
     \and
     INAF--Osservatorio di Astrofisica e Scienza dello Spazio di Bologna, Via P. Gobetti 101, I-40129 Bologna, Italy
     \and
     Cerro Tololo Inter-American Observatory/NSFs NOIRLab, Casilla 603, La Serena, Chile
     \and
     Las Campanas Observatory, Carnegie Observatories, Colina El Pino, Casilla 601, La Serena, Chile
     \and
     Scuola Normale Superiore, P.zza dei Cavalieri, I-56126 Pisa, Italy
   }

   \date{Received September 15, 1996; accepted March 16, 1997}


  \abstract
   {The identification of bright quasi-stellar objects (QSOs) is of fundamental importance to probe the intergalactic medium and address open questions in cosmology.  Several approaches have been adopted to find such sources in the currently available photometric surveys, including machine learning methods. However, the rarity of bright QSOs at high redshifts compared to other contaminating sources (such as stars and galaxies) makes the selection of reliable candidates a difficult task, especially when high completeness is required.}
   {We present a novel technique to boost recall (i.e., completeness within the considered sample) in the selection of QSOs from photometric datasets dominated by stars, galaxies, and low-$z$ QSOs (imbalanced datasets).}
   {Our heuristic method operates by iteratively removing sources whose probability of belonging to a noninteresting class exceeds a user-defined threshold, until the remaining dataset contains mainly high-$z$ QSOs.  Any existing machine learning method can be used as the underlying classifier, provided it allows for a classification probability to be estimated. We applied the method to a dataset obtained by cross-matching PanSTARRS1 (DR2), Gaia (DR3), and WISE, and identified the high-$z$~QSO candidates using both our method and its direct multi-label counterpart.}
   {We ran several tests by randomly choosing the training and test datasets, and achieved significant improvements in recall which increased from $\sim$~50\% to $\sim$~85\% for QSOs with $z>2.5$, and from $\sim$~70\% to $\sim$~90\% for QSOs with $z>3$. Also, we identified a sample of 3098 new QSO candidates on a sample of 2.6~$\times 10^6$ sources with no known classification. We obtained follow-up spectroscopy for 121 candidates, confirming 107 new QSOs with $z > 2.5$. Finally, a comparison of our QSO candidates with those selected by an independent method based on GAIA spectroscopy shows that the two samples overlap by more than 90\% and that both selection methods are potentially capable of achieving a high level of completeness. }
   {}


   \maketitle
%

\section{Introduction}
\label{sec:intro}

Light from distant and powerful quasi-stellar objects (QSO) has proven to be a useful tool to probe the inter-galactic medium \citep[IGM,][]{2009-Meiksin_PhysicsOfIGM, 2016-McQuinn_EvolutionOfIGM, 2020-Peroux_BatyonMetalCycles}, investigate fundamental physics \citep{2022-Murphy_FundamentalPhysicsESPRESSO}, carry out cosmological studies \citep{2022-Grazian_DensityOfLuminousQSOs}, probe the growth of supermassive black holes \citep{2021-Trakhtenbrot-highZSMBH}, and even probe the dynamics of the Universe \citep{2008-Liske_CosmicDynamicsELT, 2020-Boutsia_SpecFollowUp, 2023-Cristiani_GaiaZ}. The role of such cosmological beacons is expected to become even more important in the next decade with the upcoming availability of high-resolution spectrographs on the 30m-class telescopes. Therefore, comprehensive catalogs of high-$z$ bright QSOs are of the utmost importance to fulfill these goals, and a significant effort has been devoted from the astronomical community to identify new QSOs using machine learning techniques \citep[e.g., ][]{2019-BailerJones-QSOInGaia, 2019-Jin-QSOSelectionML, 2020-Wolf_UltraLumQSOs, 2021-Wenzl_RFSelection, 2021-Nakoneczny_PhotoSelection, 2023-Rodrigues-JPASQSOSelection}.  Similar techniques have been used to identify stars and extragalactic sources \citep[e.g.,][]{2019-Khramtsov, 2021-Nakazono, 2022-Barbisan, 2022-Hughes}.

\begin{figure}[!hbt]
	\includegraphics[width=\columnwidth]{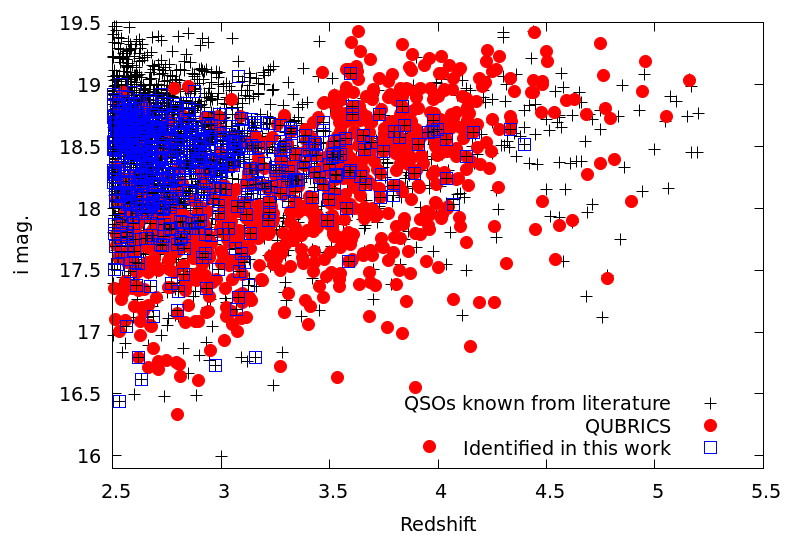}
    \caption{Known QSO with $\delta < 0$ and $i$ mag $<19.5$ from the considered catalogs in literature (black symbols, see Sect.~\ref{sec:datasets}), from the QUBRICS survey (red circles), and identified by the reverse selection method in this work (blue squares, see Sect.~\ref{sec:observations}).}
    \label{fig:ivsz}
\end{figure}

Most of the past research in the field has been, however, carried out using QSO samples in the northern hemisphere mainly due to the sky coverage of large-area surveys such as the Sloan Digital Sky Survey \citep[SDSS,][]{2020-Lyke_SDSS16Q}. However, several advanced facilities such as the Ultraviolet and Visual Echelle Spectrograph (UVES), the Echelle SPectrograph for Rocky Exoplanets and Stable Spectroscopic Observations (ESPRESSO), and the ArmazoNes high Dispersion Echelle Spectrograph (ANDES) are or will be in operation in the southern hemisphere. In order to fill this gap, we started the QUBRICS\footnote{QUasars as BRIght beacons for Cosmology in the Southern hemisphere.} survey in 2018, aiming to identify the brightest high-$z$ ($z>2.5$) QSOs in the southern hemisphere, using data available in photometric databases such as the SkyMapper, the Panoramic Survey Telescope and Rapid Response System (PanSTARRS), the dark energy survey (DES), Gaia, and the Wide-field Infrared Survey Explorer (WISE) surveys, as well as machine learning selection algorithms.  So far, we have obtained 1302 high-quality spectra\footnote{Plus 149 spectra with uncertain classification due to low S/N or too few available features for a robust classification.} of QSO candidates using several facilities (Magellan telescopes: Baade/IMACS and Clay/LDSS-3, du Pont/WFCCD,
TNG/Dolores, and NTT/EFOSC2). Among these, 1219 (94\%) were actual new Actve Galaxy Nuclei (AGN) or QSO identifications, with 943 of them (72\%) having $z > 2.5$, and 1079 (83\%) having $z > 2$. We also identified 123 QSOs with $z > 4$, with the highest redshift being 5.16.  The remaining sources were stars (56) and galaxies (27). Fig.~\ref{fig:ivsz} shows the distribution in the $i$~mag versus $z$ plane of the QSOs known from literature and of those identified by QUBRICS.
The QUBRICS survey has already produced several papers discussing both of the selection methods \citep{2019-Calderone_CCA, 2021-Guarneri_PRF} and the exploitation of new QSO identifications \citep{2020-Boutsia_SpecFollowUp, 2021-Boutsia_LuinosityFunction, 2022-Cupani_NIRSpectroscopyOfBALs, 2022-Grazian_DensityOfLuminousQSOs}.

The task of identifying bright, high-$z$ QSO candidates in photometric catalogs is the classical needle-in-a-haystack problem. For the catalogs considered in this work (Sect.~\ref{sec:datasets}), there are roughly 10$^4$ stars and 100 galaxies for each bright QSO with $i \lesssim 19$ and $z > 2.5$, that is, the dataset is imbalanced toward stars. Hence the selection methods need to be carefully tuned, and their performance constantly monitored, in order to minimize the required telescope time and maximize the success rate.  In the QUBRICS case the success rate (or precision, Sect.~\ref{sec:estimators}) has always been $\sim 70$~\% \citep{2019-Calderone_CCA, 2021-Guarneri_PRF}, and has steadily improved up to the most recent observing runs.
The latest progress has been driven by the adoption of the probabilistic random forest \citep[PRF,][]{2021-Guarneri_PRF} and XGBoost (this work) algorithms in place of the canonical correlation analysis used in the first works \citep{2019-Calderone_CCA, 2020-Boutsia_SpecFollowUp}. The initial goal of QUBRICS, namely to identify the brightest QSOs at redshift $z>2.5$ to probe the IGM and the dynamics of the Universe, has been partly achieved as shown in Fig.~\ref{fig:ivsz}.

For other cosmological studies, however, it is mandatory to achieve a high and well-determined recall (i.e., the completeness within the considered photometric sample, Sect.~\ref{sec:estimators}). The recall is more difficult to estimate than precision since the former can only be assessed indirectly by estimating the (unknown) true number of high-$z$ QSOs in the subsample still lacking spectroscopic classification.  Precision, on the other hand, can be extrapolated even with a limited set of observations. Until recently, the main obstacle has been the limited overlap between the southern hemisphere surveys used to search for new high-$z$ QSOs (namely, SkyMapper, DES, PanSTARRS) and other surveys with significantly higher QSO completeness in the North (such as SDSS). Depending on the adopted method and the underlying assumptions, we estimated a recall between 70\% and 85\% \citep{2019-Calderone_CCA, 2021-Guarneri_PRF, 2022-Guarneri-PRF-wSynth}.  With the latest observations we now have more than 900 QSOs at $z > 2.5$ and 600 at $z > 3$, that we can split into training and test datasets (with the latter containing a few tens of objects), in order to evaluate the recall in a self-consistent manner.

The QUBRICS survey aims to identify the remaining, not yet identified, QSOs in the redshift range 2.5 -- 5.0 (with the upper limit due to the requirement of having a Gaia detection).  Hence we need to maximize the recall, even though this could cause a reduced precision.
We also need to take into account the overwhelming number of noninteresting sources (mainly stars) present in the photometric datasets compared to the number of bright and high-$z$ QSOs, which results in highly imbalanced datasets (Sect.~\ref{sec:datasets}, see also Table~\ref{tab:datasets}).
The goal of this work is to present a method to boost the recall of a machine learning multi-label selection algorithm, under the only assumption that the latter provides an estimate of the probability for a source to belong to a given class.  It operates by iteratively discarding objects with high probability of not being a QSO, thus automatically rebalancing the input datasets, and providing higher recall rates with respect to other classification methods (Sect.~\ref{sec:boosting}).

The paper is organized as follows: Sect.~\ref{sec:estimators} describes the selection performance estimators used throughout the paper; Sect.~\ref{sec:boosting} describes our method to boost recall in the multi-label selection case and in Sect.~\ref{sec:dataanalysis} we apply it to the specific problem of identifying new high-$z$ QSO candidates; Sect.~\ref{sec:observations} reports the observations of such candidates, and the comparison with the QSO candidates obtained with an independent method.  Finally, in Sect.~\ref{sec:conclusions} we draw our conclusions.

\section{Performance metrics}
\label{sec:estimators}

In order to measure the performance of a selection method, we need proper metrics as described in this section.  We introduce the concepts of ``positive'' (P) and ``negative'' (N) classes in the context of a binary classification, where the former represents the class of objects of interest for a specific purpose, and the latter contains all the objects supposed to be rejected by a selection algorithm.  Whenever an object is correctly classified as belonging to either the P or N class we call it a ``true positive'' (TP) or ``true negative'' (TN) prediction respectively.  Whenever the prediction is wrong we call it a ``false negative'' (FN) or ``false positive'' (FP) respectively.  The number of ``true positives'', ``false negatives'', etc. can be arranged in a tabular form known as ``confusion matrix'' (see Table~\ref{tab:Definitions} for an example).
\begin{table}
  \centering
  \caption{Representation of a confusion matrix for a two-class (P versus N) problem. The rows represent the sources whose actual class is P or N. The columns represent the Predicted Positive (pP) and Predicted Negative (pN) sources respectively.}
  \label{tab:Definitions}
  \begin{tabular}{rccc}
    & & \multicolumn{2}{c}{\bf Predicted class:}\\
    & & {\bf pP} & {\bf pN}\\
    \cline{3-4}
    {\bf True}   & {\bf P} & TP (true positives)  & FN (false negative)\\
    {\bf class:} & {\bf N} & FP (false positive)  & TN (true negative) \\
    \cline{3-4}
  \end{tabular}
\end{table}
With these definitions, the relevant metrics are:
\begin{eqnarray}
  \nonumber
      {\rm Precision} &=& \frac{\rm TP}{\rm pP},\\
      \label{eq:Definitions}
      {\rm True\ Positive\ Rate\ (TPR\ or\ recall)} &=& \frac{\rm TP}{\rm P},\\
      \nonumber
      {\rm False\ Positive\ Rate\ (FPR)} &=& \frac{\rm FP}{\rm N}\\
      \nonumber
\end{eqnarray}
where ``pP'' represents the total number of sources predicted to belong to the positive class (i.e., the sum along the first column), while ``P'' and ``N'' represent the total number of sources actually belonging to each class (i.e., the sum along the rows).

In a classification process the ``precision'' metric is the expected success rate in identifying new sources belonging to the P class, while the ``recall'' is the expected P-class completeness within the considered sample.  We note that the above metrics may be biased if the datasets are imbalanced (e.g., when N $\gg$ P) and tend toward values which depend on the class ratio P / N rather than measuring the actual capabilities of the method.  As a consequence, there is not an absolutely ``good'' or ``bad'' value for precision and recall, since they depend on the specific case.  On the other hand, it is always possible to compare the performance of two algorithms and decide which one provides the best precision or recall performance, regardless of the ``goodness'' of the absolute values of such metrics.

If the classifier algorithm also provides a probability estimate for a source to belong to a specific class, it is possible to adopt a discriminating threshold and accept a P-classification as reliable only if its probability exceeds the threshold \citep{1997-Provost_ROC, 2008-Provost_MLFromImbalanced}.
In a single run of a binary classifier this would alter the metrics in a correlated way.  As an example, a conservative discriminating threshold would typically lead to higher precision but lower recall and FPR metrics (see Sect.~\ref{sec:binary_case} and Fig.~\ref{fig:metrics_direct}).

The above mentioned concepts can be easily extended to the multi-label case by introducing the relevant classes such as stars, galaxies, low-$z$ QSOs, high-$z$ QSOs etc., in place of the P and N ones. For instance, the top panel of Table~\ref{tab:results_direct} provides the example of a confusion matrix in the multi-label case.  An alternative representation of the same confusion matrix is given by normalizing each row by the total number of sources predicted to belong to a class (middle panel of Table~\ref{tab:results_direct}), or by the total number of sources actually belonging to each class (lower panel of Table~\ref{tab:results_direct}).  In the former case the diagonal numbers represent the precision (i.e., the expected success rate in identifying members of a given class), while in the latter case they represent the recall of the method (i.e., the completeness for a given class within the considered sample).

As mentioned in Sect.~\ref{sec:intro}, we are mainly interested in high redshift sources with $z \gtrsim 2.5$ hence we consider separate classes for low-$z$ and high-$z$ QSOs, with $z=2.5$ as discriminating threshold.  Within the high-$z$ QSO class we are much more interested in sources with $z>3$ or 4 rather than those at $z \sim 2.5$, but the latter are more abundant than the former due to the uneven redshift distribution of detectable QSOs.  As a consequence, the recall metric for high-$z$ QSO might be biased to represent the population at $z \sim 2.5$, providing little information about the selection performance at $z>3$. To overcome this issue, we introduced a new recall metric at $z>3$ by considering only the QSOs with $z>3$ in the TP and P calculations.  As we subsequently show in the next sections, the recall at $z>3$ is typically higher than the standard recall metric at $z=2.5$, the reason being that the QSOs with redshift slightly larger than $z = 2.5$ may easily be misclassified as low-$z$ QSOs resulting in a lower high-$z$ recall.

For the sake of completeness, we introduce here the Normalized Median Absolute Deviation metric (NMAD) which we use to estimate the scatter when comparing the estimated and true redshift of the QSO candidates in Sect.~\ref{sec:redshift_estimation}:
\begin{eqnarray}
  \nonumber
      {\rm NMAD} = 1.4826 \ {\rm median}(|z_{\rm est} - z_{\rm true}|).
\end{eqnarray}
The NMAD is more robust than the standard deviation in presence of outliers, and the normalization factor 1.4826 makes the two quantities equal for a normal distribution \citep{1993-Rousseeuw_NMAD, 2013-Leys_NMAD}.

\section{Boosting recall in selection methods}
\label{sec:boosting}

As discussed in Sect.~\ref{sec:intro}, the purpose of this work is to present a method to significantly boost recall in QSO selection over highly imbalanced photometric datasets (at the possible expense of slightly reducing the precision), while keeping the number of additional hyper-parameters\footnote{A parameter affecting the learning process of an algorithm.} at a minimum.  The use of imbalanced datasets in machine learning algorithms is, however, known to be detrimental to performance \citep{2009-Prati_ImbalancedData:ConceptsMethods}, as well as being a source of bias for the performance estimators themselves \citep[e.g.,][]{2004-Batista_Oversampling}.  The issue is even more compelling when trying to identify the brightest QSOs, since they represent only a small subsample of the considered source catalogs, and their identification may be challenging.  Several approaches have been suggested to address the issue of rebalancing an imbalanced dataset \citep[e.g.,][]{2004-Batista_Oversampling, 2009-Prati_ImbalancedData:ConceptsMethods}, such as ``undersampling'' (random elimination of sources in the majority class, in our case: the stars); ``oversampling'' (random duplication of sources in the minority class, in our case: high-$z$ QSOs); ``synthetic data generation'' (simulate the availability of further data for the minority class, approach discussed in \citealt{2022-Guarneri-PRF-wSynth}). These methods, although effective in rebalancing the dataset, come with drawbacks such as the possible elimination of relevant sources in the undersampling case, the possible overfit due to replication of nonrelevant features in the oversampling case, and the difficulties associated with conveying useful knowledge to the machine learning method by means of synthetic data, and generalizing the results to avoid being model-dependent.  Moreover, each of the above methods requires the addition of one or more hyper-parameters to the already long list characterizing each machine learning methods, making the exploration of the hyper-parameters space even more challenging. Besides, the worse performance of selection algorithms dealing with imbalanced training datasets may not be due to the imbalance itself, but possibly also to the ``class overlapping'' issue, that is, the difficulty in distinguishing members of two different classes with the available information \citep{2004-Prati_ImbalanceVsOverlap}. \citet{2013-Smith_InstanceHardness}, in particular, provides a definition of the instance hardness as the likelihood for a source to be misclassified, and proposes an undersampling method to rebalance the training dataset by removing problematic sources with high instance hardness (using a calibrated threshold). Such supervised undersampling method is called Instance Hardness Threshold (IHT).  Other methods based on threshold tuning to address specific problems are discussed in \citet{2016-Zou-FindBestThreshold, 2021-Johnson_RobustThresholdingForImbalancedData}. Several other methods had been proposed to deal with imbalanced datasets \citep{2019-Fernandez_LearningFromImbalanced}.

Our heuristic method is similar to the IHT, but focuses on removing sources with high probability of being noninteresting ones, rather than on those being hard to classify. It builds upon existing classifier algorithms such as random forest \citep[e.g.,][]{2018-Parmar_RandomForestReview}, probabilistic random forest \citep{2019-Reis_PRF}, gradient boosting \citep{2001-Friedman_GradientBoosting}, or any other classifier able to provide an estimate of the classification probability.  Note in particular that our method capability to handle missing values (which are very common in astronomical photometric data) is exactly the same as the underlying classifier algorithm.  The following sections illustrates the method in the binary case (Sect.~\ref{sec:binary_case}) and in the general multi-label case(Sect.~\ref{sec:multiclass_case}).

\subsection{The binary classifier case}
\label{sec:binary_case}

By varying the classification probability threshold (hereafter, $\tau$) to reliably accept a P-classification we can alter the algorithm performance metrics.  More specifically, by increasing $\tau$ the number of sources whose classification probability exceeds the threshold would be smaller, and both TP and FP decrease. As a consequence, pP decreases faster than TP, and the resulting precision is increased. On the other hand, the recall and FPR decrease since both P and N are fixed (although unknown). Fig.~\ref{fig:metrics_direct} shows an example of such correlations.

In the binary case the association of the ``interesting'' sources with the P or N class is arbitrary and we can consider the case of exchanging their roles. The TP in the first case becomes the TN in the swapped case, FP $\rightarrow$ FN and, most importantly, FPR~$\rightarrow$~$1 - {\rm TPR}$.  The consequence is that applying a threshold $\tau$ to accept a P-classification amounts to decrease its FPR, and boost the recall on the complementary class (N).

\subsection{The multi-class case (reverse selection method)}
\label{sec:multiclass_case}

The photometric sample discussed in Sect.~\ref{sec:datasets} contains at least four different classes: stars, galaxies, low-$z$ and high-$z$ QSOs.  A multi-label classifier, hereafter ``direct selection method'', trained using four such classes can be built upon a binary one following either the one-vs-rest or one-vs-one heuristics \citep{2000-Allwein_ReducingMulticlassToBinary}. In the former case we train a binary classifier to distinguish objects of one class (say high-$z$ QSOs) from all other sources, repeat for all available classes (four in our case), and consider the label with the highest score as the output classification. In the one-vs-one heuristic all possible combinations of class pairs are fed to a binary classifier, and the final prediction is based on the majority of votes in each run.  However, the performance of both heuristics is badly affected by the fact that there is no mitigation for the imbalanced datasets.

The method proposed here is a mixture of the one-vs-rest and undersampling techniques, the former being necessary to apply a threshold $\tau$ on the classification probability for a source in the P class (thus improving the recall of the sources in the N class), and the latter being used to discard all sources belonging to the noninteresting P class (thus rebalancing the datasets). We note that in this case the discarded sources are not chosen randomly as in the standard undersampling approach (Sect.~\ref{sec:boosting}), but chosen among the sources with a high probability of belonging to the noninteresting P class. Our algorithm proceeds through the following steps (see also Fig.~\ref{fig:schema}):
\begin{figure*}[!hbt]
	\includegraphics[width=\textwidth]{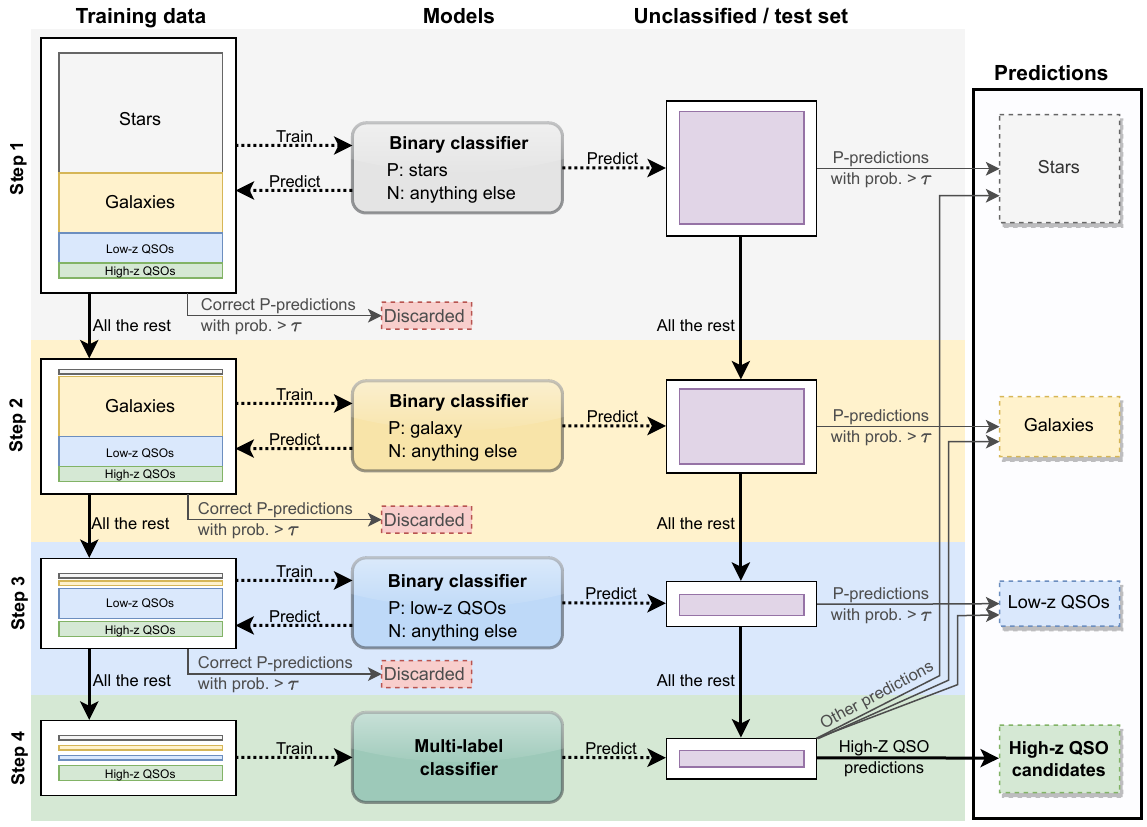}
    \caption{Schema of the reverse selection method.  In the first three steps a binary classifier is used to predict classification on all datasets, including the training one.  If the probability of belonging to the noninteresting P-class is greater than a threshold $\tau$ the source is discarded before proceeding to the next step.  By doing so all datasets decrease in size and, most importantly, they are rebalanced toward the interesting sources, namely the high-$z$ QSOs.  The last step is a simple multi-label classification, just like the direct selection method.}
    \label{fig:schema}
\end{figure*}
(i) consider the class with the largest number of sources (stars in our case).  Train a binary classifier to distinguish a star (P) from all other sources (N).  Predict a classification on the training dataset and discard all the entries with correct\footnote{We can check the predictions are correct because we are in the training dataset.  We keep the wrong predictions since they may involve the very rare high-$z$ QSOs.} P-prediction whose probability is  greater than $\tau$. Similarly, predict a classification on the test and unclassified (Sect.~\ref{sec:observations}) datasets, remove entries having  P-prediction with probability greater than $\tau$ and associate them with the ``Star'' label. Ignore predictions with probabilities $\le \tau$;  (ii) repeat for the next most abundant classes, namely galaxies and and (iii) low-$z$ QSOs; (iv) train a multi-label classifier to associate a label to the remaining sources in the test and unclassified datasets. All classifications are accepted as reliable in this step, that is, no threshold on the classification probability is adopted.
We note that, unlike other methods based on an a-posteriori ``moving threshold'' \citep[e.g.,][]{2021-Esposito_GHOST, 2021-Baqui_miniJPASStarGalaxyClassif}, our method requires the models to be re-trained whenever the threshold $\tau$ is changed.

The method sketched above is supposed to provide a higher recall metric than its direct multi-class counterpart, although with a possibly slightly lower precision.  Also, we note that the method is extremely simple to implement, and is agnostic with respect to the underlying classification framework, provided the latter allows a classification score or probability to be estimated. The interpretation of the classification probability threshold $\tau$ is straightforward: the higher the threshold, the more conservative the method is in discarding sources.  Finally, it deals naturally with highly imbalanced datasets by discarding noninteresting sources and simultaneously rebalancing all datasets (Table~\ref{tab:rebalance} shows the size the relevant datasets at each step of the method). Our approach is dubbed reverse selection method, since it focuses on the items to be removed from the photometric datasets, rather than on those to keep, in order to maximize the recall on high-$z$ QSOs.

\section{Datasets and features}
\label{sec:datasets}

The photometric dataset used throughout this paper has been prepared as follows: we selected all objects from the PanSTARRS1 (DR2) survey\footnote{\url{https://outerspace.stsci.edu/display/PANSTARRS/PS1+StackObjectView+table+fields}} \citep{2016-Chambers_PanSTARRS1}, with declination $< 15^{\circ}$, galactic latitude $>25^\circ$ (in absolute value) and Y band PSF magnitude $14 < Y < 19$.\footnote{The lower limit on magnitudes is used since our training data does not cover such bright objects at high redshifts.} We considered the PanSTARRS magnitudes in the $g$, $r$, $i$, $z$, and $Y$ bands.  Also, we cross-matched the resulting table with Gaia DR3 \citep{2022-GaiaDR3} with a matching distance of 0.5''. To avoid possibly spurious matches, we discarded all sources with multiple matching counterparts ($\sim$~0.004\% of the total).  We considered the G, RP, and BP magnitudes.  Finally, we cross-matched with the AllWISE catalog \citep{2010-AllWise} with a matching distance of 0.5'', and discarded sources with multiple matching counterparts ($\sim$~0.06\%).  We considered the magnitudes in all the four WISE bands.
The overall sample contains 30,796,027 sources (hereafter ``main sample'').  We identified stars in this sample as those sources having a proper motion or a parallax (as measured by Gaia) greater than zero with a 3$\sigma$ confidence level. Also, we cross-matched the main sample against several catalogs from the literature (\citealt{2001-Colless_2dF, 2009-Jones_6dF, 2010-Veron_Catalog}, \citealt{2016-Yang_LuminoudHighzQSOSDSS_WISE}, \citealt{2019-SchindlerPS, 2019-SchindlerSDSS, 2020-Wolf_UltraLumQSOs, 2020-Lyke_SDSS16Q, 2021-Onken_SkyMapper, 2022-Onken_UltraLumQSOSII}), as well as from our QUBRICS catalogs \citep{2019-Calderone_CCA, 2020-Boutsia_SpecFollowUp, 2021-Guarneri_PRF, 2022-Guarneri-PRF-wSynth} to assign a spectroscopic classification and, for AGN, QSOs, and galaxies, a redshift. For 2,639,184 sources we did not find any classification, hence this subset constitutes our ``unclassified'' sample where we search for new high-$z$ QSO candidates to be observed spectroscopically.

The composition of the main sample is shown in Table~\ref{tab:datasets}.
\begin{table}
	\centering
	\caption{Composition of the main sample used in this work.  The last line represents the sources for which we could not find a spectroscopic classification, nor significant proper motion or parallax measurements in the Gaia catalog, hence it is the subset we use to search for new high-$z$ ($z>2.5$) QSO candidates (Sect.~\ref{sec:observations}).}
	\label{tab:datasets}
    \begin{tabular}{|l | r | r |}
      \hline
          {\bf Class} & {\bf N. of sources} & {\bf Fraction}\\
          \hline
          Stars                  & 27,985,913  &              90.875\% \\
          Galaxies               &    150,581  &               0.489\% \\
          Low-$z$ QSOs and AGNs  &     16,269  &               0.053\% \\
          High-$z$ QSOs          &      1,908  &               0.006\% \\
          Other$^*$              &      2,172  &               0.007\% \\
          Unclassified           &  2,639,184  &               8.570\% \\
          \hline
    \end{tabular}

    {\raggedleft $^*$: any other spectral classification, such as Type 2 AGN, HII region, BL Lac, etc.}
\end{table}
As expected, the main sample turns out to be highly imbalanced toward stars and galaxies, and the high-$z$ QSOs are just ``needles in a haystack'' (0.006\% of the whole main sample). A similar order of magnitude imbalance likely affects the unclassified sample, and that a simplistic approach such as observing the whole unclassified sample would definitely recover a recall of 100\%, but would also be extremely inefficient in terms of telescope time since the maximum achievable precision would be of the order $10^{-4}$.

We used the magnitude difference between neighboring bands (i.e., colors), rather than the magnitudes themselves, as features to train the classifier model since they provide better performance when searching for very bright and rare QSOs \citep{2022-Guarneri-PRF-wSynth}. We note that identifying the optimal feature selection is not a purpose of this work: we are just interested in comparing the reverse selection method with its direct counterpart using the color features.

\section{Methodology}
\label{sec:dataanalysis}

In the following sections we apply three different selection methods to the dataset described above, and compare their precision and recall metrics in identifying high-$z$ QSOs in photometric catalogs.  We note that our analysis does not require a specific classifier algorithm, the only requirement is that it provides an estimate of the probability for a source to belong to a given class.  In this work we used XGBoost\footnote{\url{https://xgboost.ai/}} \citep{2016-Chen_XGBoost} as the underlying framework for all our analysis.

\subsection{Training and test datasets}
\label{sec:split}

In the following sections we describe two subsets of the main sample for which we have a reliable spectroscopical classification and redshift estimate into two subsets.  The first one, containing 22,525,474 sources (i.e., 80\% of the whole sample), is used to train the classifier. The second one, containing 5,631,369 sources (20\% of the sample) is used to estimate its performance and generate the confusion matrix.  We chose the 80-20 splitting schema rather than, say, 70-30 since the former is closer to the case described in Sect.~\ref{sec:observations} where we used the 100\% of the known dataset to train the models. The split is chosen randomly by shuffling the data before splitting, and following a ``stratified'' approach, that is, by keeping the class ratios approximately constant in both the training and test datasets. We did not use any validation dataset since the optimal value for the $\tau$ parameter can be established using the procedure described in Sect.~\ref{sec:probthresh}.  Also, the extreme imbalance would not allow us to have a sufficient number of high-$z$ QSO in all datasets.  We note that we used exactly the same training-test split for the specific analysis runs discussed in the following sections and to produce the results shown in Table~\ref{tab:results_comparison} and Appendix \ref{sec:confusion_matrices}.  When multiple runs are involved (Fig.~\ref{fig:metrics_direct} and \ref{fig:threshold}, Sect.~\ref{sec:montecarlo}) we generated randomized training-test splits as described above.
The unclassified dataset contains 2,639,184 sources ($\sim$~8.6\% of the main sample) and it is used in Sect.~\ref{sec:observations} to identify the list of high-$z$ QSOs candidates for the observations.

\subsection{XGBoost hyperparameters}
\label{sec:selection}

\begin{figure}[!hbt]
  \includegraphics[width=.95\columnwidth]{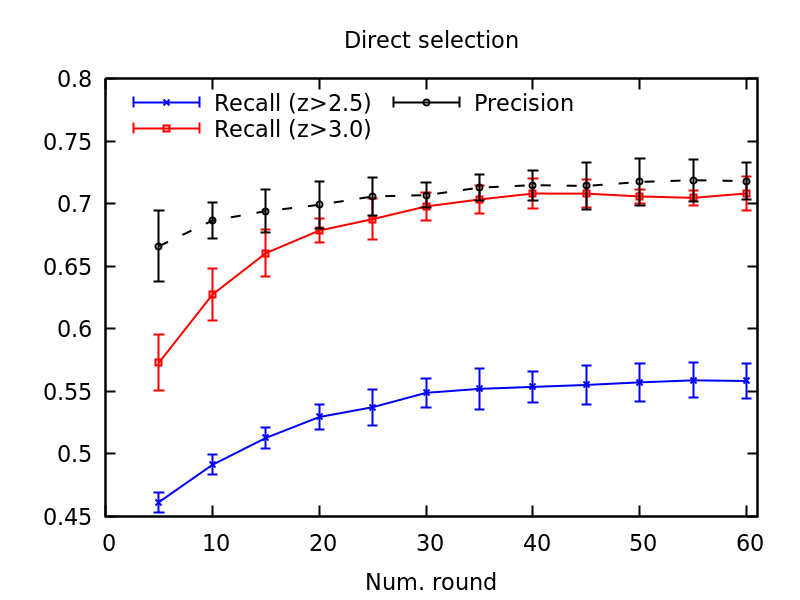}\\
  \includegraphics[width=.95\columnwidth]{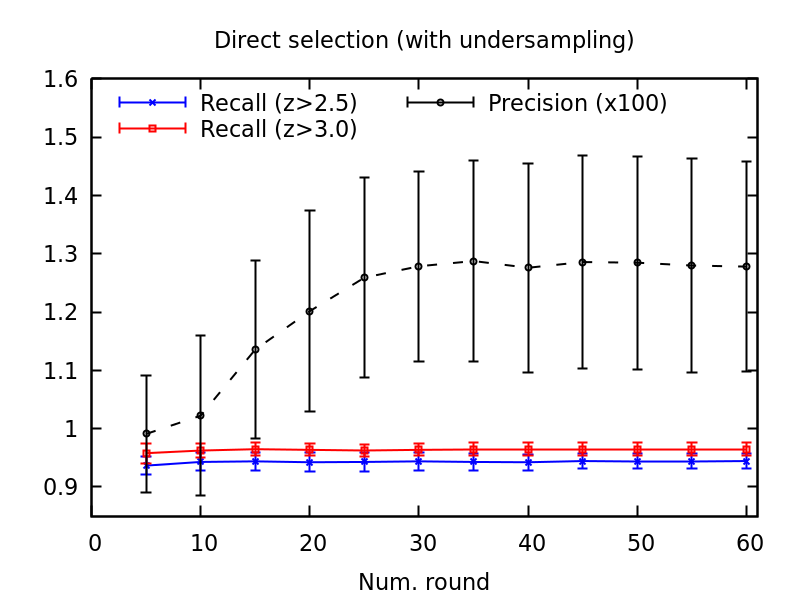}\\
  \includegraphics[width=.95\columnwidth]{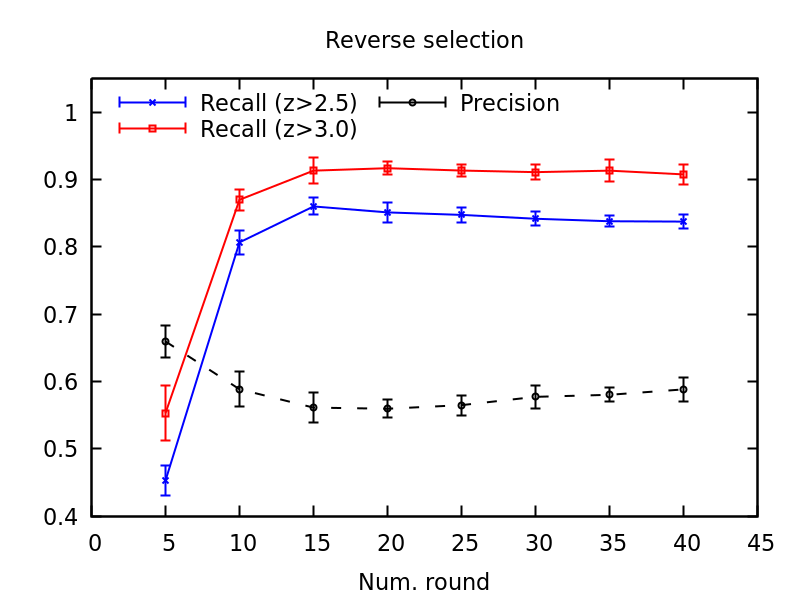}
  \caption{Precision (dashed black line), recall at $z>2.5$ (blue line), and at $z>3$ (red line) as a function of the XGBoost {\tt num\_round} parameter for the direct selection method (upper panel), direct selection with undersampling (middle panel), and reverse selection (lower panel) methods.  The precision curve in the middle panel is multiplied by a factor 100 to provide a clearer view.  The point and error bars represent respectively the mean and the standard deviation over five runs with randomly chosen training-test splits.}
    \label{fig:tuning}
\end{figure}

The values of the hyperparameters for the XGBoost classifier are as as follows:\footnote{The list of all XGBoost hyper-parameters, along with their default values, is available at \url{https://xgboost.readthedocs.io/en/stable/parameter.html}} we set \verb|num_round| (number of iterations for boosting) to 40 for the direct selection, 15 for the direct selection with undersampling, and 20 for the reverse selection method.  The maximum depth of a decision tree (\verb|max_depth|) is set to 20, and the tree construction algorithm (\verb|tree_method|) to \verb|hist| as this is the suggested setting for large datasets.  Finally, (learning objective) (\verb|objective|) is set to \verb|multi:softprob| as this is the only setting producing probability estimates for a source to belong to a class, in a multi-class problem.
The value for \verb|num_round| has been chosen as the one where the recall curves for the considered selection method stop increasing, as determined from Fig.~\ref{fig:tuning} which shows precision and recall metrics averaged over five runs (with randomly chosen training-test splits) for the direct (upper panel), direct with undersampling (middle panel) and reverse (lower panel) selection methods. We note that at higher values of \verb|num_round| both precision and recall are approximately constant, hence our results would be scarcely affected by adopting larger values.  The reverse selection method involves training several models (see Fig.~\ref{fig:schema}), hence a \verb|num_round| value may in principle be identified for each step.  However, the performance at each step is affected by the outcomes of all other steps, and identifying the optimal \verb|num_round| values for each step separately does not necessarily yield the maximum overall recall. A thorough exploration of the parameter space would thus be required to identify optimal \verb|num_round| values for each step.  This is beyond the scope of this paper, hence we simply adopt the same \verb|num_round| value for each step in the reverse selection method.

The \verb|max_depth| parameter controls the allowed complexity of the model, hence it is strictly related to {\tt num\_round}: we may achieve similar performance by increasing the former and decreasing the latter.  The chosen value of \verb|max_depth|~=~20 enables us to achieve the performance shown in Fig.~\ref{fig:tuning} in a reasonable time (training time is $\sim$~2 minutes).  All the remaining hyper-parameters are left at their default values.
We note that the probabilities provided by XGBoost are not necessarily calibrated, that is, their reliability diagrams \citep{2005-Niculescu_CalibratedProbs} may deviate from the linear 1:1 relation.  This is, however, not a problem for the reverse selection method since the probabilities are relevant only when compared to the threshold $\tau$, which in any case needs to be calibrated for the specific problem following the procedure described in Sect.~\ref{sec:probthresh}. Hence there is no need to calibrate both the probabilities and the threshold.   Finally, the method described in Sect.~\ref{sec:multiclass_case} aims to be agnostic with respect to the underlying classifier, hence the XGBoost framework used in the following sections may be replaced with any other as long as a classification probability can be estimated.

\subsection{Results with the direct selection method}
\label{sec:run_direct}

The XGBoost framework natively provides multi-label classification capabilities, hence we used it to train a model and predict the classes of the test dataset without applying any threshold on classification probability.  The relevant metrics for the test dataset of a specific run are shown in Table~\ref{tab:results_comparison}, while the entire confusion matrix is available in the appendix (Table~\ref{tab:results_direct}).  On this specific run, the direct selection method features a 70\% precision and 57\% recall for the high-$z$ QSOs.  The relatively low value for the recall is likely due to the aforementioned class imbalance problem, as shown by the significantly higher values obtained for the stars and galaxies.
%
\begin{table}
  \centering
  \caption{Comparison of the metrics for the selection of high-$z$ QSOs using the three methods discussed in this work.}
  \label{tab:results_comparison}
  \begin{tabular}{| l | c | c |}
    \hline
    {\bf Method:}                                                 &  {\bf Precision}  &  {\bf Recall} \\
    \hline
    Direct (Sect.~\ref{sec:run_direct})                           &      70.3\%       &     57.1\%    \\
    Direct with undersampling (Sect.~\ref{sec:run_undersampling}) &       1.1\%       &     94.8\%    \\
    Reverse (Sect.~\ref{sec:run_reverse})                         &      55.1\%       &     85.6\%    \\
    \hline
  \end{tabular}
\end{table}

\begin{figure}[!hbt]
	\includegraphics[width=\columnwidth]{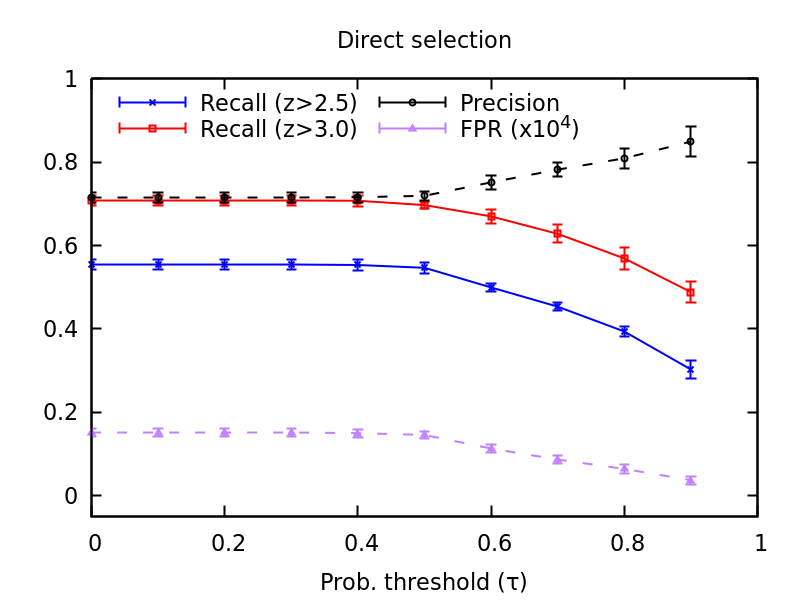}
    \caption{Precision (dashed black line), recall at $z>2.5$ (blue line), and at $z>3$ (red line) as a function of the classification probability threshold $\tau$ for the direct selection method (Sect.~\ref{sec:run_direct}) over five runs (with randomly chosen training-test splits).  The point and error bars represent respectively the mean and the standard deviation over the five runs.}
    \label{fig:metrics_direct}
\end{figure}

Although no probability threshold is involved in the direct selection method we can, a-posteriori, apply one such threshold to ignore those predictions whose confidence is too low \citep[e.g.,][]{2021-Esposito_GHOST, 2021-Baqui_miniJPASStarGalaxyClassif}. By considering such cases as noninteresting ones (i.e., not a high-$z$ QSO) we can estimate threshold-dependent precision, recall, and FPR metrics.  Fig.~\ref{fig:metrics_direct} shows such values calculated over five runs (with randomly chosen training-test splits).
As expected, applying a threshold allows the algorithm to select only the most reliable high-$z$ QSOs candidates, increasing the precision but lowering the recall and FPR metrics, in contrast to our goal of raising the recall. In principle, we can tune the threshold to achieve arbitrarily high values for the precision.  On the other hand, there is no way to tune the threshold to improve the recall, as the latter achieves the maximum value when no threshold is being applied (i.e., when $\tau=0$\footnote{For a generic multi-label classification problem with N classes all predictions have probabilities $\ge$~1/N, hence any threshold $\tau$ smaller than 1/N would be equivalent to $\tau=0$.}).

\subsection{Results with the undersampling heuristic}
\label{sec:run_undersampling}

A common approach to deal with class imbalance is to use the undersampling heuristic \citep{{2019-Fernandez_LearningFromImbalanced}}, namely to discard randomly chosen sources in the majority class(es) in order to rebalance the training dataset. We followed this approach by creating a balanced training dataset which consists of 1,526 sources (i.e., 80\% of the total number of high-$z$ QSOs, see Table~\ref{tab:datasets}) randomly chosen from each class, and used it to train a new classifier following the direct selection method to predict the classification on the test dataset. We note that we did not attempt to undersample the test dataset, which is now much larger than the training set.  The justification for this choice is that we want to use a test dataset with the same, or similar, imbalance of the unclassified dataset, which we cannot undersample or rebalance.

The relevant metrics for the test dataset of a specific run are shown in Table~\ref{tab:results_comparison}, while the entire confusion matrix is available in the appendix (Table~\ref{tab:results_undersampling}).  On this specific run, the direct selection method with under sampled training dataset provides a significant boost in recall reaching 95\%, but the precision falls down to $\sim$~1.1\% and the number of predicted high$-z$ QSOs increases to 33,207 sources (to be compared to the 310 candidates from Table~\ref{tab:results_direct}). In other words, the undersampling classifier predicts a high$-z$ QSO much more frequently than the simple direct method, achieving a very large recall rate. However, the resulting list of high-$z$ QSO candidates identified in the unclassified sample would be so large that it would be impossible to perform a spectroscopic follow-up. Since the precision is so low, we no longer consider the undersampling heuristic in the following discussion.

\subsection{Results with the reverse selection method}
\label{sec:run_reverse}

We now consider applying the reverse selection method sketched in Sect.~\ref{sec:multiclass_case} using a probability threshold of $\tau = 0.9$ (see Sect.~\ref{sec:probthresh} for a justification). In brief, the purpose of the method is to iteratively rebalance the datasets toward the high-$z$ QSOs by discarding sources with high probability (exceeding $\tau$) of being noninteresting ones. Table~\ref{tab:rebalance} shows the size of the datasets and the high-$z$ QSO fractions at each step of a specific run of the reverse selection method. We note that the size of all datasets decreases, and that the high-$z$ fractions increases significantly from $\sim$0.007\% at the beginning of first step, to $\sim$~20\% in the last step, allowing to approximately get rid of the imbalance dataset problem.  Similarly, also the number of sources in the unclassified dataset decreases, possibly increasing the fraction of high-$z$ QSO contained therein.
\begin{table}
	\centering
	\caption{Evolution of dataset sizes and class imbalance at the beginning of each step of the reverse selection method.  The first three rows report the size of the training, test, and unclassified datasets.  The last two rows report the fraction of high-$z$ QSOs in the training and test datasets (this quantities are not available for the unclassified dataset).}
	\label{tab:rebalance}
    \begin{tabular}{| l | r | r | r | r |}
      \hline
      &  {\bf Step 1} & {\bf Step 2} & {\bf Step 3} & {\bf Step 4}\\ \hline
          {N$^{\rm train}$}                 &  22,525,474 & 138,505 & 18,568 & 5,749 \\
          {N$^{\rm test}$}                 & 5,631,369  & 32,141  & 5,237  & 1,891 \\
          {N$^{\rm unclassified}$ }          & 2,639,184 & 1,324,967 & 107,193 &  81,853 \\ \hline
          {f$^{\rm train}_{\rm high-z\ QSO}$}  &    0.007\%  & 1.102\%  & 8.218\%  & 26.544\% \\
          {f$^{\rm test}_{\rm high-z\ QSO}$}   & 0.007\%  & 1.126\%  & 6.912\%  & 17.610\% \\ \hline
    \end{tabular}
\end{table}

The relevant metrics for the test dataset of a specific run are shown in Table~\ref{tab:results_comparison}, while the entire confusion matrix is available in the appendix (Table~\ref{tab:results_reverse}).  On this specific run, we obtained a significantly higher recall (85.6\%) at the cost of a relatively lower precision (55.1\%). We note that the loss in precision is not dramatic, unlike in the undersampling heuristic. On the other hand, the high recall is not due to a specific choice of the training-test split in a specific run, as shown by the point in Fig.~\ref{fig:threshold} corresponding to $\tau=0.9$, whose symbol and error bar represent the average and standard deviation over five runs (with randomly chosen training-test splits).

\subsection{Estimation of recall improvement}
\label{sec:montecarlo}

\begin{figure*}[!hbt]
	\includegraphics[width=\textwidth]{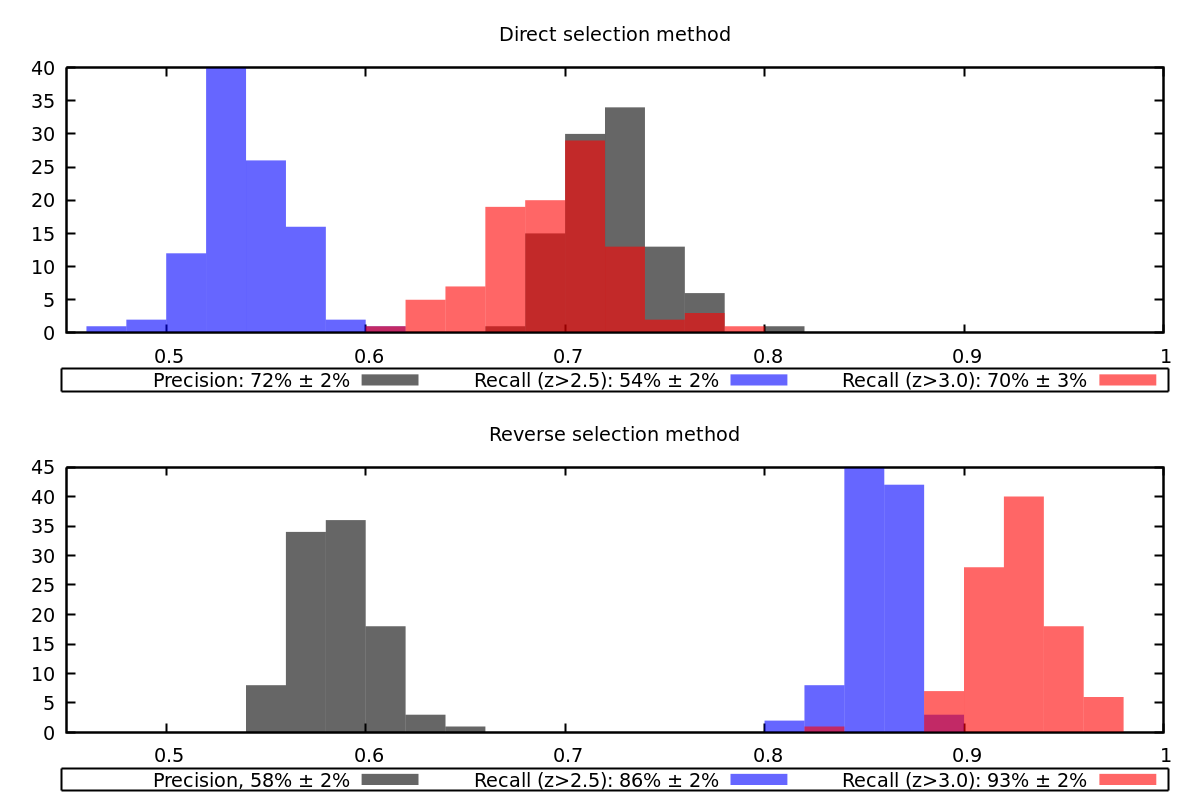}
    \caption{Distributions of precision (black histogram) and recall (blue histogram for the sample of high-$z$ QSOs with $z>2.5$, red histogram for the sample with $z>3$) as measured on 100 randomly selected test datasets (Sect.~\ref{sec:montecarlo}). The upper panel shows the results obtained with the direct selection method (Sect.~\ref{sec:run_direct}) while the lower panel shows the same quantities obtained with the reverse selection method (Sect.~\ref{sec:run_reverse}).}
    \label{fig:montecarlo}
\end{figure*}

In order to quantify the recall improvements due to the reverse selection method when compared to the direct selection one we ran 100 analysis for both the direct and reverse selection method, randomly selecting the training and test datasets at each run (following the same procedure outlined in Sect.~\ref{sec:split}). The histograms of the resulting precision and recall (for QSOs with $z>2.5$ and $z>3$) are shown in Fig.~\ref{fig:montecarlo}.
The legend also reports the mean and standard deviation for each metric, confirming that the reverse selection method allows us to improve the recall from $\sim$~50\% to $\sim$~85\%, and that the recall for the QSOs with $z>3$ is $\gtrsim$~90\%, while the precision shows only a slight decrease from $\sim$~70\% to $\sim$~60\%, when compared to the direct selection method.

\subsection{Optimal value for the probability threshold $\tau$}
\label{sec:probthresh}

The reverse selection method relies on a probability threshold $\tau$ to be calibrated in order to achieve the best possible results.  Having a threshold which is too low implies we are rejecting sources for which the classification is questionable, and this may be harmful for recall.  On the other hand, having a threshold which is too high prevents the algorithm from rebalancing the datasets, again limiting the recall.  Hence, the proper value depends on the specific problem and possibly on the underlying classifier, and should be identified by exploration of the possible range until the recall is maximized.  Note in particular that the search for the optimal values for $\tau$ implies that a calibration of the probability estimates \citep{2005-Niculescu_CalibratedProbs} is, in general, not needed.

\begin{figure}[!hbt]
	\includegraphics[width=\columnwidth]{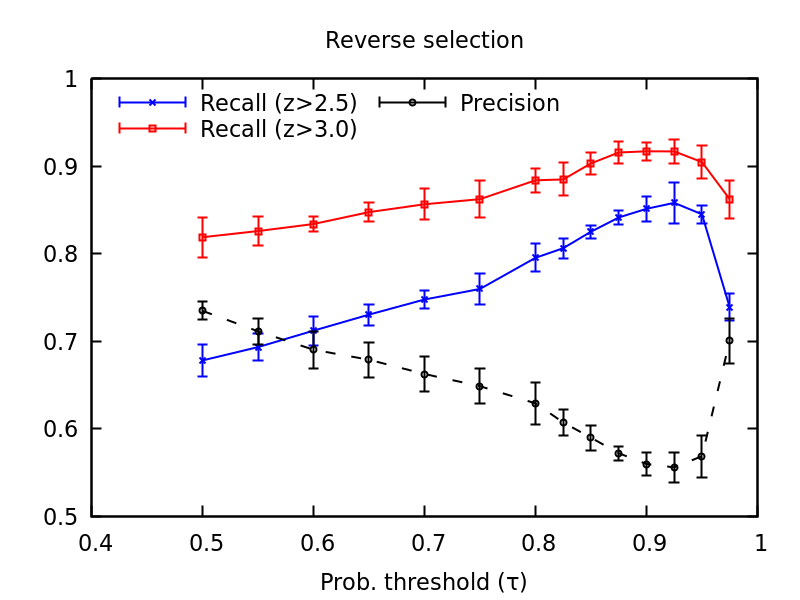}
    \caption{Precision (dashed black line) and recall as a function of the classification probability threshold $\tau$ for the reverse selection method (Sect.~\ref{sec:run_reverse}) over five runs (with randomly chosen training-test splits). The point and error bars represent respectively the mean and the standard deviation over the five runs.  The recall lines are calculated considering the whole sample of high-$z$ QSOs with $z > 2.5$ (blue line) and its subsample of QSOs with $z>3$ (red line).  The recall curves peak, and have the smaller scatter, at a value of $\tau=0.9$, hence this is the value adopted as threshold for the analysis in Sect.~\ref{sec:run_reverse}.}
    \label{fig:threshold}
\end{figure}

Fig.~\ref{fig:threshold} shows a comparison between the high-$z$ QSO precision (dashed black line) and recall (solid blue line) for the reverse selection method, as a function of the probability threshold $\tau$.  Each point represents the mean and standard deviation over five runs (with randomly chosen training-test splits).  We note that the trends of the precision and recall ($z>2.5$) curves are symmetric to those observed in Fig.~\ref{fig:metrics_direct}: as $\tau$ increases the recall is boosted only for the reverse selection method.  On the other hand, precision is somewhat reduced since in the last multi-label step the training set contains high-$z$ QSOs as well as all those sources whose classification were uncertain in the previous steps, hence they are hardly representative of the remaining classes (stars, galaxies, etc.).  The recall of the reverse selection method is maximized, and has a smaller scatter, for $\tau=0.9$, hence this is the value adopted in this work Sect.~\ref{sec:multiclass_case} and Sect.~\ref{sec:montecarlo}.

\section{Redshift estimation}
\label{sec:redshift_estimation}

Besides classification, we also need to estimate the redshift of the QSO candidates in order to prioritize the observations (i.e., giving higher priority to higher redshift candidates, Sect.~\ref{sec:observations}).  Many approaches are available to estimate a photometric redshift \citep[e.g.,][]{2021-Brescia_PhotoZ} with different level of complexity and resulting accuracy.  The requirement for the prioritization, however, is not particularly tight and any method providing redshift accuracy better than $\sim$~0.5 is enough.  Hence we simply used the regression capabilities provided by XGBoost to estimate the redshift for the low-$z$ and high-$z$ QSOs candidates identified with the reverse selection method. Specifically, we used the redshift values and the same features described in Sect.~\ref{sec:dataanalysis} to train a regression model using the XGBoost framework, with the objective set to \verb|reg:squarederror| and both \verb|num_round| and \verb|max_depth| equal to 15.  The latter values were chosen by requiring the \verb|rmse| metric (root mean square error) on the test data to reach a minimum.

\begin{figure}[!hbt]
	\includegraphics[width=\columnwidth]{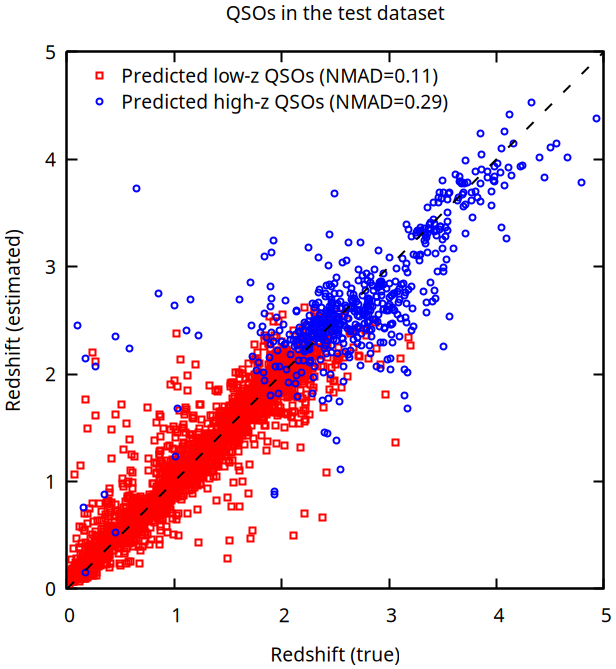}
 \caption{Comparison of estimated and true redshifts for low-$z$ and high-$z$ QSO candidates in the test dataset.}
    \label{fig:cmpz}
\end{figure}

The comparison of the true and predicted redshift values for the low-$z$ and high-$z$ QSOs candidates in the test dataset is shown in Fig.~\ref{fig:cmpz}. Only in a few cases the predicted redshift is significantly different than the true one, especially in the upper left corner where predicted high-$z$ QSOs have an estimated redshift smaller than 2.5. This suggests that the standard deviation of the difference (0.47 for the high-$z$ QSOs) may be biased by the presence of outliers.  A more robust estimator of the scatter in presence of outliers is the NMAD \citep{2013-Leys_NMAD}, whose value for the predicted low-$z$ and high-$z$ QSOs is, respectively, 0.11 and 0.29.  Among the 530 sources predicted to have $z>2.5$, we found 35 outliers ($\sim$~6\%) whose estimated redshift lies at more than 3~$\times$~NMAD from the true redshift.

\section{The high-$z$ QSO candidates sample}
\label{sec:observations}

The high-$z$ QSO candidates in this sample were selected among the unclassified data sources with no spectroscopic classification (Table~\ref{tab:datasets}) using the entire known dataset to train the models in the reverse-selection method (no test dataset is required for this purpose).  The selection method identified 3098 sources as high-$z$ QSOs candidates, corresponding to $\sim$~0.1\% of the whole unclassified sample.

We selected the targets for spectroscopic follow-up from our list of candidates according to the observing period and giving higher priority to brighter sources in the $i$ band, as well as to higher estimated redshifts (Sect.~\ref{sec:redshift_estimation}). So far, we carried out a spectroscopic observations for 121 candidates, and identified 107 new QSOs with $z$ > 2.5 (success rate of 88\%), 2 stars and 12 QSOs with $z<2.5$.  The new QSO identifications are listed in Table~\ref{tab:newqso}. Moreover, using data collected from the literature, which we ignored at the time of selection, we obtained 21 new spectroscopic redshifts (all $>$~2.5) in our unclassified sample.  In total, we could identify 143 QSOs among our candidate sample, which adds to the previously known QUBRICS QSOs (see Fig.~\ref{fig:ivsz}).

We also compared our list of candidates with the catalog of QSO redshifts obtained by a newly developed spectral energy distribution fitting technique exploiting both photometric information and Gaia DR3 spectroscopy \citep{2023-Cristiani_GaiaZ}.  Such catalog contains 1672 new redshift estimates, out of which 1142 fall in the same footprint as the dataset described in Sect.~\ref{sec:datasets}, and 717 have a redshift greater than 2.5.  We cross-matched this 717 sources with our candidates and found 658 sources in common (92\%). If we consider only the sources with estimated redshift greater than 3, our candidates sample includes 103 of the 110 sources also selected by Cristiani et al. (94\%). Such high overlaps between QSO candidates identified with independent methods (although on similar initial samples) show that both are potentially able to reach high level of completeness.  Our method, however, relies on photometric estimates and does not require the Gaia spectra.

The Cristiani et al. sample also includes 1142 - 717 = 425 sources with $z <= 2.5$, corresponding to 381 sources in our sample.  These have to be considered ``failures'' of our selection method since their redshift is smaller than 2.5. They are present because of the limitations of our algorithm in correctly classifying QSOs whose redshift is close to the threshold of $z=2.5$. In fact, our redshift estimates for such spurious source is in the range $2 \lesssim z_{\rm XGBoost} \lesssim 3$, with only 2 sources having an estimated redshift greater than $z=3$. As already mentioned in Sect.~\ref{sec:probthresh}, this issue is the reason to introduce the recall metrics at $z>3$ (rather than $z>2.5$) to estimate the performance of our method. Concerning the other metrics, the presence of these spurious sources with $z<2.5$, besides the above mentioned 658 ones with $z>2.5$, in our candidate sample implies an upper limit of 658/(658+381) = 63\% for the precision of our method, which is in line with the precision estimated with the analysis described in Sect.~\ref{sec:montecarlo}. We note that we may improve the precision by simply neglecting the sources with estimated redshift smaller than 2.5.  In this case, we would approximately halve the candidate sample (1563 sources rather than 3098), but the recall with respect to the Cristiani et al. sample of QSOs with $z>3$ would fall to $\sim$~78\%. The joint adoption of multiple selection algorithms, such as reverse selection and SED fitting \citep{2023-Cristiani_GaiaZ} may improve the redshift estimates and the precision while keeping a recall $\sim$~90\%. This will be the subject of a future work.

\section{Conclusions}
\label{sec:conclusions}

We presented a novel heuristic method, dubbed reverse selection method, designed to improve the recall (i.e., the completeness over the considered dataset) of a classifier algorithm, even in the presence of a highly imbalanced dataset, at the expense of a slight decrease in precision. The method relies on the adoption of a classification probability threshold for the validation of the outcome of a binary classifier, in order to improve its precision.  When applied repeatedly following the class size order (i.e., starting from stars, then galaxies, low-$z$ QSOs, etc.) this allows us to identify and remove noninteresting objects in order to rebalance the datasets toward the less common sources (high-$z$ QSOs).

We applied the reverse selection method to search for high-$z$ QSOs in a highly imbalanced dataset where most of the sources are stars or galaxies, and compared the precision and recall to its simple, direct selection multi-label classifier counterpart, both with and without random undersampling.  Our results confirm that the reverse selection method provides a significant boost in recall, up to 90\% (for QSOs with $z > 3$) with only a small decrease in precision ($\sim$~60\% rather than $\sim$~70\%).  In order to show that the improvement in recall is not due to a particular split of the datasets, we tested the robustness of our results by randomly generating the training and test datasets at each run, confirming that our method is capable of achieving a recall of $\sim$~90\% (for QSOs with $z>3$).

Our heuristic method relies on an external classifier algorithm, which is XGBoost for the analysis discussed here.  However, the method is agnostic with respect to the underlying classifier, and an alternative algorithm providing a classification probability estimate can in principle be used. The method relies on the probability threshold $\tau$ whose interpretation is straightforward and whose optimal value for a specific case can be easily tuned. Also, the boost in recall metrics can be easily quantified following the procedure outlined in Sect.~\ref{sec:montecarlo}.

Finally, we applied our method to a sample of objects without any known spectroscopic classification, and identified a sample of 3098 new QSO candidates among them. For 121 candidates we obtained a follow-up spectroscopy, and identified 107 new QSOs with $z > 2.5$.  A comparison with the recently released catalog of \citet{2023-Cristiani_GaiaZ} shows that both selection methods are able to achieve similar recall rates up to $\sim$~90\%, but our method shows such performance with no need for the Gaia spectroscopic data. On the other hand, our method still needs to be complemented with a reliable redshift estimate algorithm in order to be able to reduce the number of spurious sources while keeping the same recall rate.

\begin{acknowledgements}

We thank the anonymous Referee for the insightful comments which helped us improving the manuscript.

We acknowledge financial contribution from the grant PRIN INAF 2019 (RIC) 1.05.01.85.09: ``New light on the Intergalactic Medium (NewIGM)''.

A.G. and F.F. acknowledge support from PRIN MIUR project ``Black Hole winds and the Baryon Life Cycle of Galaxies: the stone-guest at the galaxy evolution supper'', contract
2017-PH3WAT.

A.G. acknowledges the support of the INAF Mini Grant 2022 ``Learning Machine Learning techniques to dig up high-z AGN in the Rubin-LSST Survey''.

We thank Societ\`a Astronomica Italiana (SAIt), Ennio Poretti, Gloria Andreuzzi and Marco Pedani for the observation support at TNG. Part of the observations discussed in this work are based on observations made with the Italian Telescopio Nazionale Galileo (TNG) operated on the island of La Palma by the Fundacion Galileo Galilei of the INAF (Istituto Nazionale di Astrofisica) at the Spanish Observatorio del Roque de los Muchachos of the Instituto de Astrofisica de Canarias.

This paper includes data gathered with the 6.5 meter Magellan Telescopes located at Las Campanas Observatory, Chile.

This work has made use of data from the European Space Agency (ESA) mission Gaia (https://www.cosmos.esa.int/gaia), processed by the Gaia Data Processing and Analysis Consortium (DPAC, https://www.cosmos.esa.int/web/gaia/dpac/consortium). Funding for the DPAC has been provided by national institutions, in particular the institutions participating in the Gaia Multilateral Agreement.

This publication makes use of data products from the Wide-field Infrared Survey Explorer, which is a joint project of the University of California, Los Angeles, and the Jet Propulsion Laboratory/California Institute of Technology, funded by the National Aeronautics and Space Administration.

The Pan-STARRS1 Surveys (PS1) and the PS1 public science archive have been made possible through contributions by the Institute for Astronomy, the University of Hawaii, the Pan-STARRS Project Office, the Max-Planck Society and its participating institutes, the Max Planck Institute for Astronomy, Heidelberg and the Max Planck Institute for Extraterrestrial Physics, Garching, The Johns Hopkins University, Durham University, the University of Edinburgh, the Queen's University Belfast, the Harvard-Smithsonian Center for Astrophysics, the Las Cumbres Observatory Global Telescope Network Incorporated, the National Central University of Taiwan, the Space Telescope Science Institute, the National Aeronautics and Space Administration under Grant No. NNX08AR22G issued through the Planetary Science Division of the NASA Science Mission Directorate, the National Science Foundation Grant No. AST-1238877, the University of Maryland, Eotvos Lorand University (ELTE), the Los Alamos National Laboratory, and the Gordon and Betty Moore Foundation.

We acknowledge the support from the LBT-Italian Coordination Facility for the execution of observations, data distribution and reduction. The LBT is an international collaboration among institutions in the United States, Italy and Germany. The LBT Corporation partners are: The University of Arizona on behalf of the Arizona university system; Istituto Nazionale di Astrofisica, Italy;  LBT Beteiligungsgesellschaft, Germany, representing the Max Planck Society, the Astrophysical Institute Potsdam, and Heidelberg University; The Ohio State University; The Research Corporation, on behalf of The University of Notre Dame, University of Minnesota and University of Virginia.

\end{acknowledgements}

\begin{appendix}

\section{Confusion matrices}
\label{sec:confusion_matrices}

This section reports the confusion matrices for three specific runs of the direct selection (Sect.~\ref{sec:run_direct}), direct selection with undersampling (Sect.~\ref{sec:run_undersampling}), and the reverse selection (Sect.~\ref{sec:run_reverse}) methods. In all cases we used exactly the same training-test split.  Relevant metrics highlighted in bold are (from top to bottom): the true positives (TP), the number of predicted positives (pP), the precision and the recall at $z > 2.5$.  The latter two are also reported in Table~\ref{tab:results_comparison} for an easy comparison of the performance of the three methods.

\begin{table*}
	\centering
	\caption{Confusion matrix, precision, and recall for the direct selection method (Sect.~\ref{sec:run_direct}).}
	\label{tab:results_direct}
\begin{tabular}{| l  l | r | r | r | r | r | r |}
\hline
&  & \multicolumn{5}{c|}{\bf Predicted class:} & \\
& {\bf Confusion matrix:} &           Galaxy &         QSOhighZ &          QSOlowZ &             Star &            other &            TOTAL \\ \cline{2-8}
\multirow{5}{*}{\rotatebox[origin=c]{90}{\bf True class:}}
&                 Galaxy &           26399  &               1  &             282  &            3428  &               6  &            30116 \\
&               QSOhighZ &               0  & {\bf        218} &             129  &              35  &               0  &              382 \\
&                QSOlowZ &             516  &              82  &            2571  &              79  &               6  &             3254 \\
&                   Star &              84  &               8  &              22  &         5597069  &               0  &          5597183 \\
&                  other &             408  &               1  &              15  &               5  &               5  &              434 \\
\cline{2-8}
&                  TOTAL &            27407 &              310 &             3019 &          5600616 &               17 &          5631369 \\ \hline\hline
&       {\bf Precision:} &           Galaxy &         QSOhighZ &          QSOlowZ &             Star &            other \\ \cline{2-7}
\multirow{5}{*}{\rotatebox[origin=c]{90}{\bf True class:}}
&                 Galaxy &           96.3\%  &            0.3\%  &            9.3\%  &            0.1\%  &           35.3\%  \\
&               QSOhighZ &            0.0\%  & {\bf      70.3\%} &            4.3\%  &            0.0\%  &            0.0\%  \\
&                QSOlowZ &            1.9\%  &           26.5\%  &           85.2\%  &            0.0\%  &           35.3\%  \\
&                   Star &            0.3\%  &            2.6\%  &            0.7\%  &           99.9\%  &            0.0\%  \\
&                  other &            1.5\%  &            0.3\%  &            0.5\%  &            0.0\%  &           29.4\%  \\
\cline{2-7}
&                  TOTAL &            100\% &            100\% &            100\% &            100\% &            100\% \\ \hline\hline
&          {\bf Recall:} &           Galaxy &         QSOhighZ &          QSOlowZ &             Star &            other &            TOTAL \\ \cline{2-8}
\multirow{5}{*}{\rotatebox[origin=c]{90}{\bf True class:}}
&                 Galaxy &           87.7\%  &            0.0\%  &            0.9\%  &           11.4\%  &            0.0\%  &            100\% \\
&               QSOhighZ &            0.0\%  & {\bf      57.1\%} &           33.8\%  &            9.2\%  &            0.0\%  &            100\% \\
&                QSOlowZ &           15.9\%  &            2.5\%  &           79.0\%  &            2.4\%  &            0.2\%  &            100\% \\
&                   Star &            0.0\%  &            0.0\%  &            0.0\%  &          100.0\%  &            0.0\%  &            100\% \\
&                  other &           94.0\%  &            0.2\%  &            3.5\%  &            1.2\%  &            1.2\%  &            100\% \\
\cline{2-8}
\hline
\end{tabular}
\end{table*}

\begin{table*}
	\centering
	\caption{Confusion matrix, precision, and recall for the direct selection method with undersampling (Sect.~\ref{sec:run_undersampling}).}
	\label{tab:results_undersampling}
\begin{tabular}{| l  l | r | r | r | r | r | r |}
\hline
&  & \multicolumn{5}{c|}{\bf Predicted class:} & \\
& {\bf Confusion matrix:} &           Galaxy &         QSOhighZ &          QSOlowZ &             Star &            other &            TOTAL \\ \cline{2-8}
\multirow{5}{*}{\rotatebox[origin=c]{90}{\bf True class:}}
&                 Galaxy &           21622  &              41  &             793  &             807  &            6853  &            30116 \\
&               QSOhighZ &               1  & {\bf        362} &              13  &               5  &               1  &              382 \\
&                QSOlowZ &              67  &             346  &            2598  &              18  &             225  &             3254 \\
&                   Star &          183999  &           32456  &            8528  &         5371001  &            1199  &          5597183 \\
&                  other &             113  &               2  &              18  &               0  &             301  &              434 \\
\cline{2-8}
&                  TOTAL &           205802 &            33207 &            11950 &          5371831 &             8579 &          5631369 \\ \hline\hline
&       {\bf Precision:} &           Galaxy &         QSOhighZ &          QSOlowZ &             Star &            other \\ \cline{2-7}
\multirow{5}{*}{\rotatebox[origin=c]{90}{\bf True class:}}
&                 Galaxy &           10.5\%  &            0.1\%  &            6.6\%  &            0.0\%  &           79.9\%  \\
&               QSOhighZ &            0.0\%  & {\bf       1.1\%} &            0.1\%  &            0.0\%  &            0.0\%  \\
&                QSOlowZ &            0.0\%  &            1.0\%  &           21.7\%  &            0.0\%  &            2.6\%  \\
&                   Star &           89.4\%  &           97.7\%  &           71.4\%  &          100.0\%  &           14.0\%  \\
&                  other &            0.1\%  &            0.0\%  &            0.2\%  &            0.0\%  &            3.5\%  \\
\cline{2-7}
&                  TOTAL &            100\% &            100\% &            100\% &            100\% &            100\% \\ \hline\hline
&          {\bf Recall:} &           Galaxy &         QSOhighZ &          QSOlowZ &             Star &            other &            TOTAL \\ \cline{2-8}
\multirow{5}{*}{\rotatebox[origin=c]{90}{\bf True class:}}
&                 Galaxy &           71.8\%  &            0.1\%  &            2.6\%  &            2.7\%  &           22.8\%  &            100\% \\
&               QSOhighZ &            0.3\%  & {\bf      94.8\%} &            3.4\%  &            1.3\%  &            0.3\%  &            100\% \\
&                QSOlowZ &            2.1\%  &           10.6\%  &           79.8\%  &            0.6\%  &            6.9\%  &            100\% \\
&                   Star &            3.3\%  &            0.6\%  &            0.2\%  &           96.0\%  &            0.0\%  &            100\% \\
&                  other &           26.0\%  &            0.5\%  &            4.1\%  &            0.0\%  &           69.4\%  &            100\% \\
\cline{2-8}
\hline
\end{tabular}
\end{table*}

\begin{table*}
	\centering
	\caption{Confusion matrix, precision, and recall for the reverse selection method (Sect.~\ref{sec:run_reverse}).}
	\label{tab:results_reverse}
\begin{tabular}{| l  l | r | r | r | r | r | r |}
\hline
&  & \multicolumn{5}{c|}{\bf Predicted class:} & \\
& {\bf Confusion matrix:} &           Galaxy &         QSOhighZ &          QSOlowZ &             Star &            other &            TOTAL \\ \cline{2-8}
\multirow{5}{*}{\rotatebox[origin=c]{90}{\bf True class:}}
&                 Galaxy &           25628  &               8  &             657  &            3389  &             434  &            30116 \\
&               QSOhighZ &               0  & {\bf        327} &              29  &              24  &               2  &              382 \\
&                QSOlowZ &             277  &             203  &            2615  &              77  &              82  &             3254 \\
&                   Star &             772  &              54  &              31  &         5596316  &              10  &          5597183 \\
&                  other &             322  &               1  &              18  &               2  &              91  &              434 \\
\cline{2-8}
&                  TOTAL &            26999 &              593 &             3350 &          5599808 &              619 &          5631369 \\ \hline\hline
&       {\bf Precision:} &           Galaxy &         QSOhighZ &          QSOlowZ &             Star &            other \\ \cline{2-7}
\multirow{5}{*}{\rotatebox[origin=c]{90}{\bf True class:}}
&                 Galaxy &           94.9\%  &            1.3\%  &           19.6\%  &            0.1\%  &           70.1\%  \\
&               QSOhighZ &            0.0\%  & {\bf      55.1\%} &            0.9\%  &            0.0\%  &            0.3\%  \\
&                QSOlowZ &            1.0\%  &           34.2\%  &           78.1\%  &            0.0\%  &           13.2\%  \\
&                   Star &            2.9\%  &            9.1\%  &            0.9\%  &           99.9\%  &            1.6\%  \\
&                  other &            1.2\%  &            0.2\%  &            0.5\%  &            0.0\%  &           14.7\%  \\
\cline{2-7}
&                  TOTAL &            100\% &            100\% &            100\% &            100\% &            100\% \\ \hline\hline
&          {\bf Recall:} &           Galaxy &         QSOhighZ &          QSOlowZ &             Star &            other &            TOTAL \\ \cline{2-8}
\multirow{5}{*}{\rotatebox[origin=c]{90}{\bf True class:}}
&                 Galaxy &           85.1\%  &            0.0\%  &            2.2\%  &           11.3\%  &            1.4\%  &            100\% \\
&               QSOhighZ &            0.0\%  & {\bf      85.6\%} &            7.6\%  &            6.3\%  &            0.5\%  &            100\% \\
&                QSOlowZ &            8.5\%  &            6.2\%  &           80.4\%  &            2.4\%  &            2.5\%  &            100\% \\
&                   Star &            0.0\%  &            0.0\%  &            0.0\%  &          100.0\%  &            0.0\%  &            100\% \\
&                  other &           74.2\%  &            0.2\%  &            4.1\%  &            0.5\%  &           21.0\%  &            100\% \\
\cline{2-8}
\hline
\end{tabular}
\end{table*}

\section{List of newly classified high-$z$ QSO}
The 107 new QSO classifications and redshift identified with the reverse selection method are reported in Table~\ref{tab:newqso}.

\onecolumn
\begin{longtable}[c]{| l | c | c | c | c | l | }
	\caption{QSO identified with the reverse selection method and observed within the QUBRICS survey.}\label{tab:newqso}\\
      \hline
          {\bf \#ID\footnote{Qubrics internal ID.}} & {\bf R.A.} & {\bf Decl.} & {\bf Mag. i}      & {\bf Redshift}  & {\bf Instrument, Obs. date} \\
                                                        &            &             & {\bf (PanSTARRS)} &                 &                              \\
          \hline
     1136419  &   00:01:21.87  &  -17:03:24.77  &  18.21  &  2.56  &  TNG, 08/2022 \\ 
     7865345  &   00:06:42.07  &  -21:11:14.00  &  18.24  &  3.35  &  LDSS-3, 11/2022 \\ 
    42578500  &   00:09:30.56  &  -25:33:13.29  &  18.39  &  3.53  &  NTT, 07/2023 \\ 
     8344752  &   00:14:13.05  &  +11:08:27.67  &  18.26  &  4.00  &  TNG, 08/2022 \\ 
     8292898  &   00:15:05.68  &  +07:47:54.50  &  18.36  &  3.55  &  TNG, 08/2022 \\ 
     1352951\footnote{Blazar candidate in \citealt{2014-DAbrusco}.}  &   00:18:29.95  &  +03:19:03.28  &  16.33  &  2.78  &  TNG, 08/2022 \\ 
     1137257  &   00:20:19.08  &  -07:39:04.20  &  18.67  &  3.85  &  LBT, 10/2022 \\ 
     1185897  &   00:20:42.57  &  -27:51:25.76  &  18.05  &  3.11  &  LDSS-3, 11/2022 \\ 
     7763709  &   00:23:07.71  &  -29:54:29.33  &  18.00  &  3.58  &  LDSS-3, 11/2022 \\ 
     1131398  &   00:24:41.06  &  -09:53:45.46  &  18.32  &  3.13  &  TNG, 08/2022 \\ 
     8044729  &   00:26:24.70  &  -08:23:00.94  &  18.23  &  3.34  &  TNG, 08/2022 \\ 
     1160487\footnote{QSO candidate in \citealt{2009-Richards}.}  &   00:33:44.11  &  -08:58:39.52  &  18.33  &  3.38  &  TNG, 08/2022 \\ 
     1186253  &   00:37:23.11  &  -12:03:00.77  &  18.21  &  3.31  &  TNG, 08/2022 \\ 
    28750884  &   00:44:01.31  &  -12:43:15.28  &  17.28  &  3.08  &  TNG, 08/2022 \\ 
     1185975  &   00:45:29.27  &  -14:27:53.73  &  18.16  &  3.52  &  TNG, 08/2022 \\ 
    28697003  &   00:49:58.97  &  -29:50:47.75  &  19.09  &  3.59  &  LDSS-3, 10/2022 \\ 
    28939580\footnote{QSO candidate in \citealt{2007-Atlee}.}  &   01:00:34.09  &  -16:24:18.19  &  18.50  &  3.14  &  TNG, 08/2022 \\ 
     1189989  &   01:02:11.44  &  -13:46:08.75  &  18.76  &  3.73  &  TNG, 01/2023 \\ 
    29100306  &   01:04:16.20  &  +06:42:59.08  &  18.65  &  3.87  &  TNG, 08/2022 \\ 
    42578520\footnote{Blazar candidate in \citealt{2019-DAbrusco}.}  &   01:15:55.64  &  -06:12:53.31  &  18.80  &  3.61  &  IMACS, 07/2023 \\ 
    20032352  &   01:26:44.53  &  +02:24:49.39  &  18.38  &  4.19  &  TNG, 08/2022 \\ 
    19922839  &   01:28:23.43  &  -05:02:09.81  &  18.54  &  3.69  &  LBT, 10/2022 \\ 
    19667617  &   01:41:41.03  &  -25:24:48.45  &  18.47  &  3.41  &  LDSS-3, 11/2022 \\ 
    19753168  &   01:47:48.56  &  -18:08:34.96  &  18.08  &  2.96  &  TNG, 01/2023 \\ 
     1190673  &   01:49:09.73  &  -22:27:32.15  &  18.19  &  3.16  &  LDSS-3, 11/2022 \\ 
    20126803  &   01:53:16.59  &  +08:51:21.54  &  18.87  &  3.67  &  TNG, 08/2022 \\ 
     1409684  &   02:00:30.15  &  -22:50:54.97  &  17.57  &  3.59  &  LDSS-3, 11/2022 \\ 
    29928553  &   02:23:25.90  &  +13:14:14.93  &  18.11  &  3.78  &  TNG, 11/2022 \\ 
    29908150  &   02:24:31.50  &  +11:01:25.84  &  18.18  &  3.39  &  TNG, 08/2022 \\ 
     1191423  &   02:27:19.70  &  -29:08:52.90  &  18.04  &  3.22  &  LDSS-3, 11/2022 \\ 
     1115350  &   02:52:19.85  &  -27:29:57.05  &  17.35  &  2.93  &  NTT, 11/2022 \\ 
    30232805  &   02:53:05.15  &  +10:06:04.77  &  18.35  &  3.15  &  TNG, 01/2023 \\ 
    30235907  &   02:57:58.30  &  +10:45:43.21  &  18.76  &  3.27  &  TNG, 11/2022 \\ 
    30313276  &   03:03:47.48  &  +11:46:48.56  &  18.55  &  3.34  &  TNG, 11/2022 \\ 
     1180142  &   03:13:29.53  &  +01:31:38.43  &  18.75  &  3.86  &  LBT, 10/2022 \\ 
    42578490  &   03:32:45.24  &  -01:58:22.40  &  18.25  &  3.35  &  NTT, 07/2023 \\ 
      853375  &   04:03:45.72  &  -21:17:35.99  &  16.79  &  2.62  &  NTT, 11/2022 \\ 
     1176698  &   04:13:13.06  &  -02:21:42.05  &  18.27  &  3.53  &  LBT, 10/2022 \\ 
    32386534  &   04:22:53.81  &  +00:25:08.13  &  18.41  &  3.46  &  TNG, 11/2022 \\ 
    33324780  &   04:45:33.58  &  -00:29:59.65  &  18.00  &  2.53  &  TNG, 01/2023 \\ 
     1179324  &   04:48:34.19  &  -23:28:25.70  &  18.54  &  3.52  &  LDSS-3, 11/2022 \\ 
     1200765  &   05:02:34.45  &  -07:30:01.37  &  18.18  &  2.79  &  TNG, 01/2023 \\ 
    33131342  &   05:06:17.43  &  -08:15:41.29  &  18.08  &  2.82  &  TNG, 01/2023 \\ 
    33053044  &   05:07:49.64  &  -10:47:05.43  &  18.05  &  2.57  &  TNG, 01/2023 \\ 
     1012567  &   05:26:56.36  &  -18:50:12.90  &  16.73  &  2.97  &  TNG, 01/2023 \\ 
     1143201  &   05:42:03.95  &  -25:19:37.78  &  18.29  &  3.43  &  LDSS-3, 11/2022 \\ 
     3353506\footnote{Also identified by \citealt{2023-Jin}.}  &   08:57:20.19  &  +13:36:46.56  &  18.15  &  2.97  &  TNG, 01/2023 \\ 
     3188759\footnote{Also identified by \citealt{2023-Jin}.}  &   09:00:48.74  &  +07:33:50.13  &  17.90  &  2.86  &  TNG, 01/2023 \\ 
      843636  &   09:11:05.22  &  -03:09:17.85  &  17.98  &  2.59  &  TNG, 01/2023 \\ 
     1142138  &   09:16:34.40  &  -05:08:14.26  &  18.06  &  2.51  &  TNG, 01/2023 \\ 
     1161386  &   09:58:53.77  &  -21:14:57.32  &  18.64  &  4.00  &  LDSS-3, 01/2023 \\ 
     5631792  &   10:37:44.75  &  -23:07:28.66  &  18.52  &  3.91  &  IMACS, 04/2023 \\ 
     5946831  &   10:41:33.54  &  -20:33:14.76  &  17.57  &  3.11  &  LDSS-3, 01/2023 \\ 
     1153144  &   10:42:24.08  &  -26:03:18.18  &  18.27  &  3.53  &  LDSS-3, 01/2023 \\ 
    42578510  &   11:27:00.11  &  -09:21:14.60  &  18.19  &  3.79  &  NTT, 07/2023 \\ 
     7452193  &   11:53:11.28  &  -10:56:38.17  &  18.38  &  2.96  &  IMACS, 04/2023 \\ 
     7280871  &   11:59:10.59  &  -24:27:09.28  &  18.76  &  3.61  &  LDSS-3, 05/2023 \\ 
    42578418  &   12:03:11.51  &  -21:59:08.36  &  18.17  &  2.64  &  LDSS-3, 05/2023 \\ 
    42578508  &   12:08:46.01  &  -12:52:54.21  &  17.92  &  3.49  &  NTT, 07/2023 \\ 
     8599727  &   12:13:57.31  &  -19:10:32.34  &  18.00  &  2.73  &  NTT, 07/2023 \\ 
     1172859  &   13:08:37.23  &  -29:55:16.67  &  18.23  &  3.32  &  LDSS-3, 01/2023 \\ 
    42578493  &   14:21:56.18  &  +09:24:45.61  &  18.66  &  3.57  &  NTT, 07/2023 \\ 
    42578515  &   14:22:43.71  &  -14:10:14.58  &  18.80  &  3.98  &  IMACS, 07/2023 \\ 
    42578516  &   14:22:48.76  &  -12:46:41.35  &  18.49  &  4.04  &  IMACS, 07/2023 \\ 
    42578491  &   14:26:40.11  &  +02:10:29.29  &  18.55  &  4.38  &  NTT, 07/2023 \\ 
    42578461  &   14:35:11.87  &  -25:40:18.69  &  18.24  &  3.64  &  IMACS, 04/2023 \\ 
    42578518  &   14:48:36.33  &  -05:50:02.41  &  18.88  &  3.84  &  IMACS, 07/2023 \\ 
    42578464  &   15:32:32.56  &  -14:33:45.09  &  18.58  &  3.82  &  IMACS, 04/2023 \\ 
    18333242  &   16:35:08.90  &  +03:11:37.26  &  18.46  &  4.17  &  TNG, 08/2022 \\ 
    18534393  &   16:46:02.46  &  +09:43:56.82  &  18.55  &  3.94  &  TNG, 08/2022 \\ 
      839089  &   20:33:15.57  &  -09:30:14.20  &  17.83  &  2.89  &  TNG, 08/2022 \\ 
    21042742  &   20:34:12.05  &  -23:31:44.81  &  18.24  &  4.04  &  LDSS-3, 10/2022 \\ 
    21720915  &   20:34:22.67  &  -05:09:20.87  &  18.64  &  4.33  &  TNG, 08/2022 \\ 
     1204030  &   20:44:55.03  &  -28:42:24.43  &  18.09  &  3.42  &  LDSS-3, 10/2022 \\ 
    42578410\footnote{QSO candidate in \citealt{2016-Khorunzhev}.}  &   20:46:02.49  &  -02:38:48.69  &  18.38  &  3.51  &  LDSS-3, 05/2023 \\ 
    22149902  &   20:48:57.06  &  -02:51:38.29  &  16.79  &  3.16  &  TNG, 08/2022 \\ 
     1145726  &   20:51:24.47  &  -03:44:45.09  &  18.42  &  4.13  &  TNG, 08/2022 \\ 
     1170286  &   20:52:15.36  &  +00:14:55.67  &  18.24  &  3.55  &  TNG, 08/2022 \\ 
     1145766  &   20:58:59.32  &  -02:57:41.74  &  18.61  &  3.96  &  TNG, 08/2022 \\ 
     1146008  &   21:01:29.92  &  -14:40:40.29  &  18.59  &  3.65  &  TNG, 08/2022 \\ 
    23914910  &   21:11:06.80  &  +06:07:53.45  &  18.43  &  3.56  &  TNG, 08/2022 \\ 
    23647497  &   21:18:40.67  &  -13:19:28.25  &  18.62  &  3.80  &  TNG, 08/2022 \\ 
     1146176  &   21:25:13.59  &  -04:41:46.75  &  18.38  &  3.28  &  LBT, 10/2022 \\ 
     1146286  &   21:33:21.26  &  -10:59:36.35  &  18.43  &  3.55  &  TNG, 08/2022 \\ 
     1146586  &   21:38:44.10  &  -15:55:39.23  &  18.30  &  3.31  &  TNG, 08/2022 \\ 
     1146966  &   21:48:06.33  &  -16:34:33.38  &  18.35  &  3.06  &  TNG, 08/2022 \\ 
    25022141  &   21:48:27.46  &  -20:56:25.97  &  18.08  &  3.73  &  LDSS-3, 10/2022 \\ 
    42578509  &   21:50:45.08  &  -10:23:04.47  &  18.55  &  4.17  &  NTT, 07/2023 \\ 
     1146979  &   21:51:05.17  &  -09:54:31.61  &  18.28  &  3.74  &  TNG, 08/2022 \\ 
    42578492  &   21:51:34.60  &  +04:44:04.24  &  18.24  &  3.55  &  NTT, 07/2023 \\ 
     1154840  &   22:01:00.08  &  -24:23:59.89  &  18.07  &  3.49  &  LDSS-3, 10/2022 \\ 
     1383551  &   22:02:32.64  &  +06:08:42.63  &  17.10  &  2.68  &  TNG, 08/2022 \\ 
     1146870  &   22:16:25.08  &  -05:54:05.28  &  18.28  &  3.89  &  TNG, 08/2022 \\ 
     1146202  &   22:16:28.93  &  -06:20:37.85  &  18.48  &  3.66  &  LBT, 10/2022 \\ 
    26965711\footnote{Blazar candidate in \citealt{2019-DAbrusco}.}  &   22:22:26.21  &  -05:29:14.78  &  17.90  &  3.22  &  TNG, 08/2022 \\ 
     1182367  &   22:23:42.06  &  -11:41:42.28  &  18.19  &  3.47  &  TNG, 08/2022 \\ 
     1156260  &   22:27:42.41  &  -18:55:08.54  &  18.03  &  4.07  &  TNG, 08/2022 \\ 
     1184176  &   22:33:12.48  &  -15:09:29.42  &  18.29  &  3.59  &  TNG, 08/2022 \\ 
    26841627  &   22:35:30.26  &  -16:25:13.42  &  18.25  &  3.45  &  TNG, 08/2022 \\ 
    27544394  &   22:44:15.37  &  +04:24:18.60  &  18.76  &  3.96  &  TNG, 08/2022 \\ 
    27352735  &   22:52:43.96  &  -06:55:02.60  &  18.56  &  3.57  &  TNG, 08/2022 \\ 
    42578494  &   23:16:47.15  &  +11:28:19.10  &  18.36  &  4.15  &  NTT, 07/2023 \\ 
     1184927  &   23:19:07.27  &  -18:59:36.62  &  18.16  &  3.20  &  TNG, 08/2022 \\ 
     1187685  &   23:20:50.47  &  -28:57:05.16  &  18.19  &  3.66  &  LDSS-3, 10/2022 \\ 
    28132326  &   23:23:31.53  &  +02:47:14.66  &  18.45  &  3.67  &  TNG, 08/2022 \\ 
    42578506  &   23:48:19.88  &  -16:54:44.15  &  18.43  &  3.75  &  NTT, 07/2023 \\ 
    42578498  &   23:50:53.91  &  -26:03:46.32  &  18.60  &  3.98  &  NTT, 07/2023 \\ 
\hline
\end{longtable}

\end{appendix}

\end{document}